\documentclass[10pt,prb,aps,superscriptaddress,twocolumn,longbibliography]{revtex4-1}
\usepackage{amssymb,amsmath}
\usepackage{graphicx}
\usepackage{color}
\usepackage{natbib}
\usepackage{hyperref}
\usepackage[utf8]{inputenc}

\begin{document}

\newcommand{\e}{{\rm e}}
\newcommand{\norm}[1]{\left\lVert#1\right\rVert}
\newcommand{\rmi}{{\rm i}}
\renewcommand{\Im}{\mathop\mathrm{Im}\nolimits}
\newcommand{\red}[1]{{\color{red}#1}}
\newcommand{\blue}[1]{{\color{blue}#1}}

\renewcommand{\cite}[1]{[\onlinecite{#1}]}

\newcommand{\ket}[1]{\left|#1\right>}      
\newcommand{\bra}[1]{\left<#1\right|}
\newcommand{\eps}{\varepsilon}      
\newcommand{\om}{\omega}      
\newcommand{\kap}{\varkappa}      

\newcommand{\skvv}[2]{\left<#1\left|#2\right.\right>} 
\newcommand{\commentMaxim}[1]{{\color{red}{\it ~Maxim:~}\tt #1}}
\newcommand{\commentLeshaG}[1]{{\color{blue}{\it ~LeshaG:~}\tt #1}}
\renewcommand{\thefootnote}{\roman{footnote}}

\title{
Photonic Jackiw-Rebbi states in all-dielectric structures controlled by bianisotropy}

\author{Alexey~A.~Gorlach}
\affiliation{Belarusian State University, Minsk 220030, Belarus}
\email{alexey.gorlach@gmail.com.  A.G. and D.Z. contributed equally to this work}

\author{Dmitry~V.~Zhirihin}
\affiliation{ITMO University, Saint Petersburg 197101, Russia}

\author{Alexey~P.~Slobozhanyuk}
\affiliation{ITMO University, Saint Petersburg 197101, Russia}
\affiliation{Nonlinear Physics Centre, Research School of Physics and Engineering, Australian National University, Canberra ACT 2601, Australia}

\author{Alexander~B.~Khanikaev}
\affiliation{ITMO University, Saint Petersburg 197101, Russia}
\affiliation{The City College of the City University of New York, New York 10031, USA}
\affiliation{Graduate Center of the City University of New York, New York 10016, USA}

\author{Maxim~A.~Gorlach}
\email{m.gorlach@metalab.ifmo.ru}
\affiliation{ITMO University, Saint Petersburg 197101, Russia}

\begin{abstract}
Electric and magnetic resonances of dielectric particles have recently uncovered a range of exciting applications in steering of light at the nanoscale. Breaking of particle inversion symmetry further modifies its electromagnetic response giving rise to bianisotropy known also as magneto-electric coupling. Recent studies suggest the crucial role of magneto-electric coupling in realization of photonic topological metamaterials. To further unmask this fundamental link, we design and test experimentally one-dimensional array composed of dielectric particles with overlapping electric and magnetic resonances and broken mirror symmetry. Flipping over half of the meta-atoms in the array, we observe the emergence of  interface states providing photonic realization of the celebrated Jackiw-Rebbi model. We trace the origin of these states to the fact that local modification of particle bianisotropic response affects its effective coupling with the neighboring meta-atoms which provides a promising avenue to engineer topological states of light.
\end{abstract}

\maketitle


{\it Introduction}~--~Electric and magnetic resonances supported by dielectric particles~\cite{Evlyukhin,Kuznetsov} have initiated an entire direction of all-dielectric nanophotonics opening a route towards metamaterials and metadevices without lossy plasmonic components~\cite{Jahani,Kuznetsov-S} featuring unique functionalities both in linear~\cite{Krasnok} and nonlinear regimes~\cite{Shcherbakov}. Bianisotropy, which stems from the particle symmetry reduction, further broadens the plethora of available physical effects enabling the coupling between electric and magnetic responses of a particle.

An important application of bianistropic meta-atoms is associated with engineering of time-reversal-invariant photonic topological metamaterials~\cite{Khanikaev,SlobNP}. In contrast to the majority of systems, photonic topological structures support electromagnetic modes confined to the boundaries and possessing a unique property of spin-locked unidirectional propagation. Besides that, photonic topological states feature a considerable robustness against disorder or imperfections which is of paramount importance for applications~\cite{Lu2014,Lu2016,Khanikaev-NP,Ozawa_RMP}. Recently, experimental works have demonstrated photonic topological structures in two~\cite{Slob16,Slob19} and three~\cite{Yang} spatial dimensions with topological edge states facilitated by magneto-electric coupling. 

\begin{center}
    \begin{figure}[b]
    \includegraphics[width=0.9\linewidth]{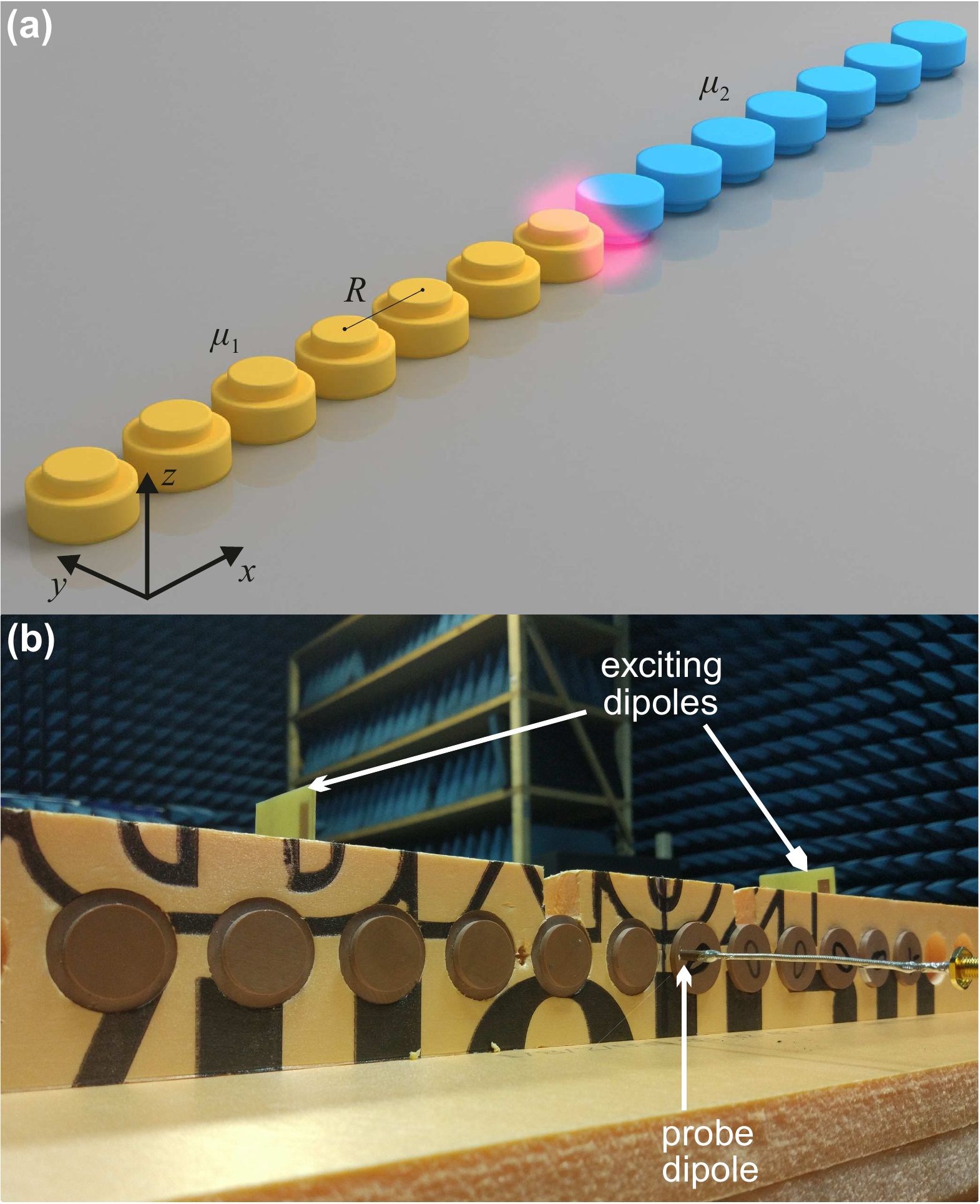}
    \caption{One-dimensional array composed of disks with different magnitudes of bianisotropy parameter: $\mu_1$ and $\mu_2$. (a) A schematic of the structure. (b) Photograph of experimental sample composed of $N=12$ disks.} 
    \label{fig:Array}
    \end{figure}
    \end{center}

The next conceptual step, in our opinion, is related to the analysis of photonic structures with spatially varying bianisotropic response which may give rise to the novel types of topologically protected states. In this Communication, aiming to further unmask the fundamental link between bianisotropy and topological states of light, we investigate an array composed of dielectric meta-atoms with broken mirror symmetry and different bianisotropic response [Fig.~\ref{fig:Array}(a)]. Based on the discrete dipole approximation, we derive an effective photonic Hamiltonian for such system and demonstrate the interface states at the boundary between the two halves of the array with different sign of magneto-electric coupling. Such states are proved to be photonic analogues of Jackiw-Rebbi states which are close relatives of topological states.

The key result of our analysis is that local modification of the disk bianisotropy translates into the modification of coupling constants in the effective photonic Hamiltonian. Apparently, this opens a route to engineer electromagnetic topological states via the staggered bianisotropy pattern. To confirm our findings, we realize such one-dimensional array experimentally [Fig.~\ref{fig:Array}(b)] proving the emergence of the interface states by mapping the near fields of the structure.

{\it Theoretical model}~--~To grasp the physics of the proposed 1D system, we develop a theoretical model based on the discrete dipole approximation~\cite{Purcell,Draine}. To simplify the treatment, we take into account only $x$ and $y$ components of electric and magnetic dipole moments of the disks neglecting off-resonant $z$-components along with higher-order multipoles. Additionally, we consider interaction of the disk with its nearest neighbors only. This yields the set of equations
\begin{equation}\label{DiscreteDipole}
\begin{split}
& \begin{pmatrix}
\hat{\alpha}^{\rm{ee}} & \hat{\alpha}^{\rm{em}}\\
\hat{\alpha}^{\rm{me}} & \hat{\alpha}^{\rm{mm}}
\end{pmatrix}
^{-1}\,
\begin{pmatrix}
{\bf p}_n\\
{\bf m}_n
\end{pmatrix}
\\
=&\sum\limits_{l=n\pm 1}\,
\begin{pmatrix}
\hat{G}^{\rm{ee}}({\bf r}_{nl}) & \hat{G}^{\rm{em}}({\bf r}_{nl})\\
\hat{G}^{\rm{me}}({\bf r}_{nl}) & \hat{G}^{\rm{mm}}({\bf r}_{nl})
\end{pmatrix}
\,
\begin{pmatrix}
{\bf p}_l\\
{\bf m}_l
\end{pmatrix}\:.
\end{split}
\end{equation}

Here, $\hat{G}({\bf r})$ is the dyadic Green's function~\cite{Novotny}; ${\bf p}$ and ${\bf m}$ are the two-component vectors containing $x$ and $y$ components of electric and magnetic dipole moments, respectively; $\hat{\alpha}^{\rm{ee}}$, $\hat{\alpha}^{\rm{mm}}$ are tensors of electric and magnetic polarizability of the disk, and off-diagonal tensors $\hat{\alpha}^{\rm{em}}$ and $\hat{\alpha}^{\rm{me}}$ describe the effect of magneto-electric coupling.

Rotational symmetry of the disk with respect to $Oz$ axis and mirror symmetry in $Oxz$ plane impose constraints on the structure of the polarizability tensors: $\hat{\alpha}^{\rm{ee}}$ and $\hat{\alpha}^{\rm{mm}}$ are proportional to the identity matrix, while the tensors of magneto-electric coupling read
\begin{equation}\label{MagnetoElectric}
\hat{\alpha}^{\rm{em}}=\hat{\alpha}^{\rm{me}}=
\begin{pmatrix}
0 & i\,\chi\\
-i\,\chi & 0
\end{pmatrix}
\:.
\end{equation}
Moreover, by the proper choice of the disks parameters we can ensure equal magnitudes of electric and magnetic polarizabilities in the frequency range of interest: $\hat{\alpha}^{\rm{ee}}=\hat{\alpha}^{\rm{mm}}=\beta\,\hat{I}$. Further details on the analysis of this model are provided in Supplementary Materials. 

Inverted polarizability tensor takes the form
\begin{equation}
\begin{pmatrix}
u & 0 & 0 & -i\,v \\
0 & u & i\,v & 0 \\
0 & -i\,v & u & 0 \\
i\,v & 0 & 0 & u
\end{pmatrix}
\:,
\end{equation}
where $u=\beta/\left(\beta^2-\chi^2\right)$ and $v=\chi/\left(\beta^2-\chi^2\right)$. Additionally, in the vicinity of the disk resonance $\omega_0$ we approximate the frequency dependence of $u$ as $\left(\omega-\omega_0\right)/A$ which corresponds to the pole approximation of polarizability. With these assumptions, Eq.~\eqref{DiscreteDipole} can be reformulated as an eigenvalue problem with the effective Hamiltonian and the ``wave function'':
\begin{equation}\label{Eigenvalue}
\hat{H}\,\ket{\psi}=\eps\,\ket{\psi}\:.
\end{equation}
Here, the wavefunction is constructed from the components of electric and magnetic dipole moments
\begin{equation}\label{Eigenvalue2}
\ket{\psi}=\left(p_x+i\,m_y, p_x-i\,m_y, p_y-i\,m_x, p_y+i\,m_x\right)^T
\end{equation}
and normalized to unity as usual, the eigenvalue is related to the frequency of the mode $\eps=2\left(\om-\om_0\right)\,R^3/A$, whereas the effective photonic Hamiltonian takes the form
\begin{equation}\label{HamiltonianFull}
\hat{H}=
\begin{pmatrix}
\hat{H}^{(+)}_{\rm{s}} & 0 \\
0 & \hat{H}^{(-)}_{\rm{s}}
\end{pmatrix}
\:,
\end{equation}
where each of the $2\times 2$ blocks is given by the expression
\begin{equation}\label{SpinBlock}
\hat{H}^{(\pm)}_{\rm{s}}=2\,\hat{I}\,\cos\,k + \mu\,\sigma_z\pm 6\,\sigma_x\,\cos\,k\:,
\end{equation}
$\mu=2 v\,R^3$ and $\sigma_i$ stand for Pauli matrices. We interpret two blocks of the Hamiltonian as two effective {\it pseudo-spins}, referred to as just a spin in the following. Spin-up configuration $\ket{\psi_+}$ corresponds to the nonzero $p_x$ and $m_y$ components of dipole moments, whereas spin-down configuration $\ket{\psi_-}$ is characterized by nonzero $p_y$ and $m_x$. Essentially, energy spectra for two pseudospins coincide.

   \begin{center}
    \begin{figure}[b]
    \includegraphics[width=0.95\linewidth]{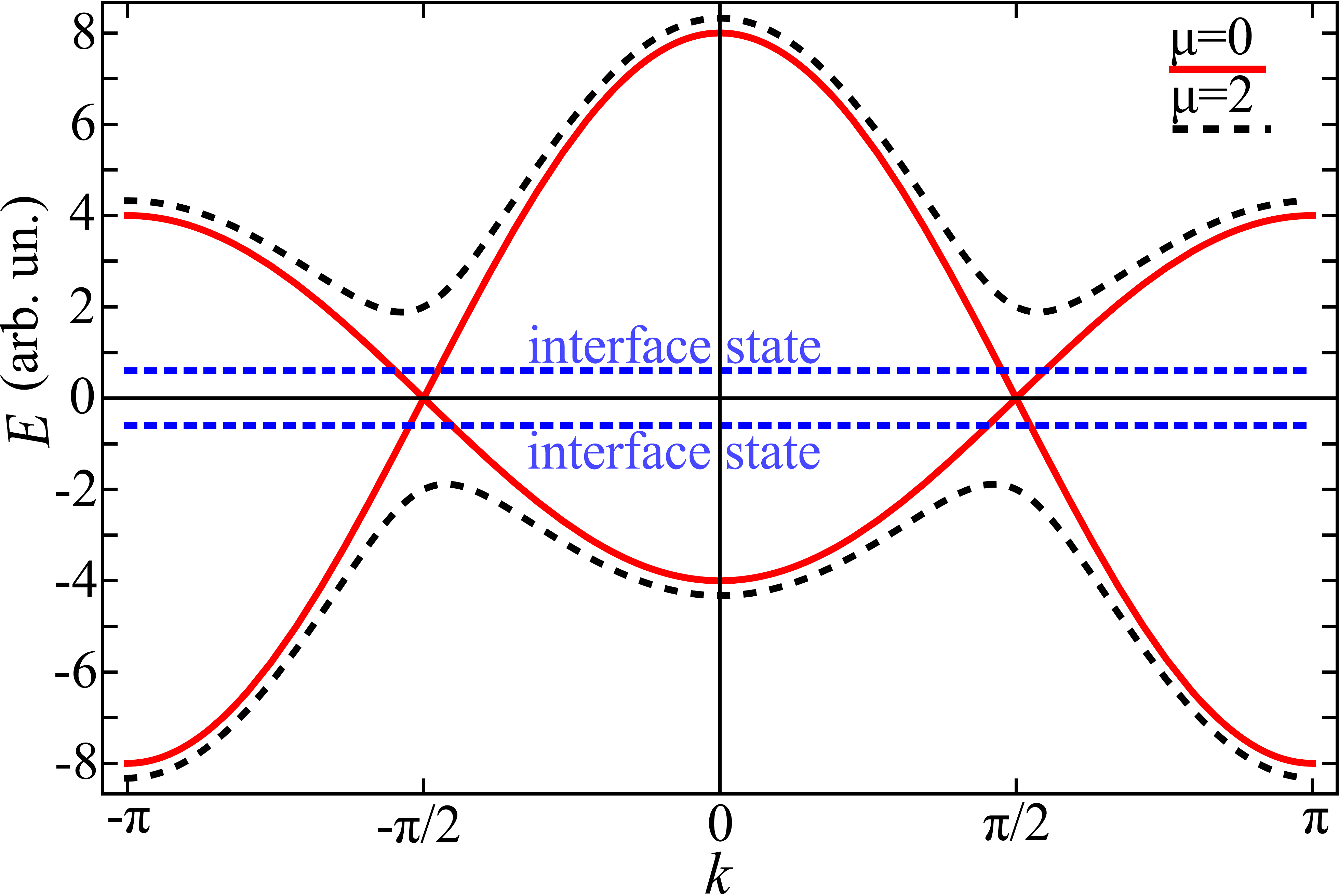}
    \caption{The dispersion relation $E(k)$ for the infinite array without and with bianisotropy (red solid  and black dashed lines, respectively). There are two Dirac-like points at $k_{1,2}=\pm\pi/2$. Dashed blue lines indicate the energy of the interface states which appear in the presence of the domain wall.}
    \label{fig:Spectrum}
    \end{figure}
    \end{center}

   \begin{center}
    \begin{figure}[b]
    \includegraphics[width=0.95\linewidth]{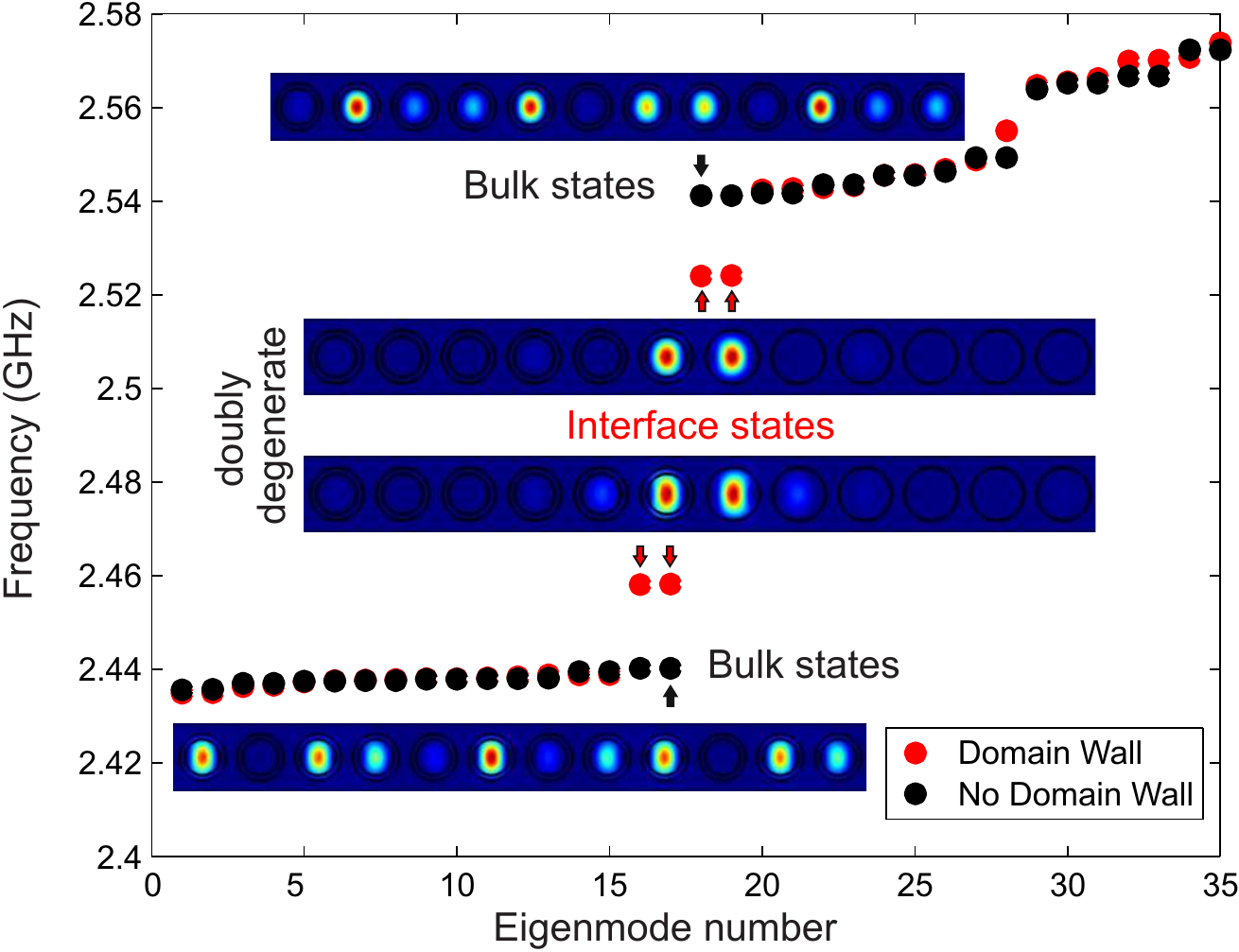}
    \caption{Eigenmodes of the array composed of ceramic disks with permittivity $\eps=39$ calculated with Comsol Multiphysics software. In numerical simulation, the array is placed in a cylindrical waveguide of the radius $r_w=21.3$~mm with perfectly conducting walls. Periodic boundary conditions along $x$ axis are applied. One period contains 16 particles. Large disk radius and height are $14.55$~mm and $9.00$~mm; small disk radius and height are $11.00$~mm and $3.00$~mm, respectively. Distance between the neighboring disks is $R=36$~mm. Insets show the calculated distribution of magnetic field $|H_x|^2+|H_y|^2$ for bulk and interface modes. Note that interaction of the array with the waveguide causes a small blueshift of the eigenmodes in comparison with the experiment performed in an anechoic chamber.} 
    \label{fig:Eigenmode}
    \end{figure}
    \end{center}

Energy spectrum calculated for the model Eqs.~\eqref{HamiltonianFull}, \eqref{SpinBlock} is shown in Fig.~\ref{fig:Spectrum}. We observe two Dirac-like points in the vicinity of $k_{1,2}=\mp\pi/2$. Expanding the Hamiltonian Eq.~\eqref{SpinBlock} in the vicinity of $k=k_{1,2}+\delta\,k$, we get
\begin{equation}\label{SpinValleyBlock}
\hat{H}_{\rm{s}}^{(\pm)}=2\,\delta k\,\tau_z+\mu\,\sigma_z\pm 6\,\sigma_x\,\tau_z\,\delta\,k\:.
\end{equation}
where the mass term $\mu=2v\,R^3$ quantifies magneto-electric coupling in the disk, and elements of Pauli matrix $\tau_z$ label inequivalent points in reciprocal space: $\tau_z=1$ for $k_1=-\pi/2$ and $\tau_z=-1$ for $k_2=\pi/2$.

Note that Eq.~\eqref{SpinValleyBlock} corresponds to the Dirac Hamiltonian~\cite{Shen}. An important feature of the Dirac equation is the emergence of zero-energy Jackiw-Rebbi states at the boundary of the domains with the opposite sign of the mass term~\cite{Shen}. 

Therefore, we conclude that the boundary between the two halves of the array with the opposite sign of magneto-electric coupling hosts pseudospin-degenerate interface states. However, due to nonlinear dependence of the original Hamiltonian on wave number $k$ [Eq.~\eqref{SpinBlock}] the  energy of these states is nonzero. On the other hand, since the system Hamiltonian possesses generalized chiral symmetry, any state with energy $+E$ is necessarily accompanied by another state with the energy $4\,\cos k-E$. Hence, taking into account double degeneracy of all states due to polarization degree of freedom, we expect two doublets of interface states, as discussed further in Supplementary Materials.

In the limit of weak magneto-electric coupling $\mu\ll 1$, the energies of the interface states can be approximated by $E=\pm\mu^2/6$, whereas for larger $\mu$ the interface states are pushed away from the bandgap center. However, they do not disappear in the continuum of bulk states even for extremely large values of magneto-electric coupling $\mu\rightarrow\infty$.

The results of presented analytical model exhibit good qualitative agreement with full-wave numerical simulations carried on in Comsol Multiphysics (Fig.~\ref{fig:Eigenmode}). The simulations reveal in particular that the spectrum of bulk excitations of the array has a bandgap, which hosts four states inside it. These four states are grouped into two doublets positioned symmetrically with respect to the bandgap center. The calculated mode profiles, shown as insets in Fig.~\ref{fig:Eigenmode}, suggest that the field for these four states is indeed localized at the domain wall.

\begin{figure}[b]
    \includegraphics[width=0.90\linewidth]{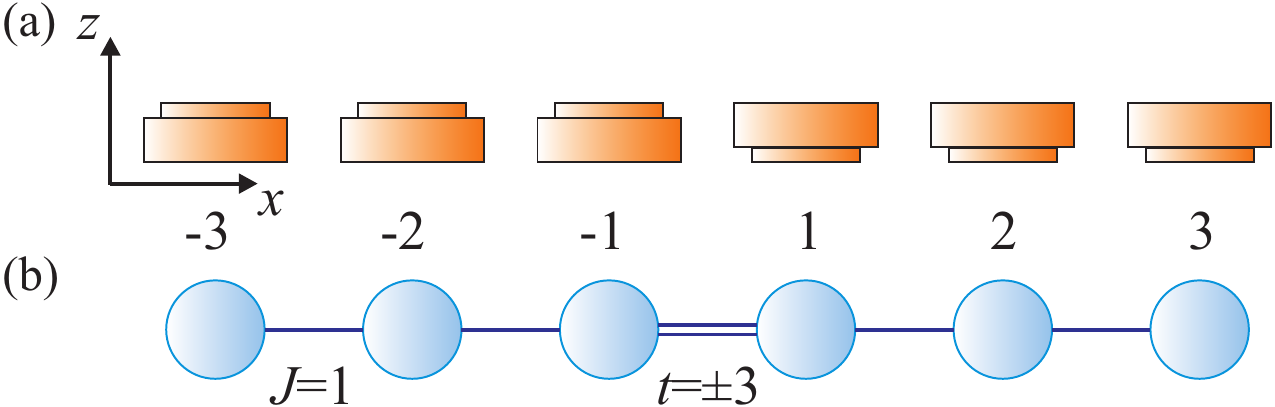}
    \caption{Mapping of electromagnetic problem (a) onto the tight-binding model (b) valid in the limit of strong bianisotropy $\mu\gg 1$. Note that local bianisotropy of meta-atoms controls the magnitude of couplings in the effective tight-binding model.} 
    \label{fig:Domain}
    \end{figure}

{\it Origin of the interface states: modification of coupling due to bianisotropy}~--~To further clarify the origin of the interface states in our model, it is instructive to examine the limit of strong bianisotropy $\mu\gg 1$. We  analyze geometry Fig.~\ref{fig:Domain}(a), in which case periodicity of array is broken by the domain wall. Therefore, Eqs.~\eqref{HamiltonianFull}, \eqref{SpinBlock} have to be replaced by
\begin{equation}\label{DomainWall}
\begin{split}
\eps\,\ket{\varphi_n}=
\begin{pmatrix}
\mu_n & 0 \\
0 & -\mu_n
\end{pmatrix}\,\ket{\varphi_n}\\
+
\begin{pmatrix}
1 & t \\
t & 1
\end{pmatrix}\,\left(\ket{\varphi_{n-1}}+\ket{\varphi_{n+1}}\right)\:,
\end{split}
\end{equation}
where the same basis Eq.~\eqref{Eigenvalue2} is used, and $\ket{\varphi}=(a_n, b_n)^T$ denotes a two-component wave function for the single pseudospin. $t=\pm 3$ for up and down pseudospins, respectively.

Now consider a fixed band with energy $\eps\approx \mu$. In our geometry, $\mu_n=\mu$ for $n<0$ and $\mu_n=-\mu$ for $n>0$ [Fig.~\ref{fig:Domain}(a)]. Then, Eq.~\eqref{DomainWall} suggests that the amplitudes $b_n$ for $n<0$ and $a_n$ for $n>0$ scale as $1/\mu$ and can be neglected in the limit $\mu\gg 1$. Truncating the system Eq.~\eqref{DomainWall} to include only the leading-order amplitudes $a_n$ for $n<0$ and $b_n$ for $n>0$, we get:
\begin{gather}
\left(\eps-\mu\right)\,a_n=a_{n-1}+a_{n+1}\:,\mspace{8mu} (n\leq -2)\label{TightBinding1}\\
\left(\eps-\mu\right)\,a_{-1}=a_{-2}+t\,b_{1}\:,\\
\left(\eps-\mu\right)\,b_{1}=t\,a_{-1}+b_{2}\:,\\
\left(\eps-\mu\right)\,b_{n}=b_{n-1}+b_{n+1}\:.\mspace{8mu} (n\geq 2)\label{TightBinding2}
\end{gather}
In turn, the system of equations \eqref{TightBinding1}-\eqref{TightBinding2} corresponds precisely to the tight-binding model Fig.~\ref{fig:Domain}(b) which provides a simple interpretation of predicted interface states. Most importantly, this derivation clearly demonstrates that the magnitude of coupling which enters the effective photonic Hamiltonian is controlled by the signs of bianisotropy of the two adjacent disks and can be enhanced up to three times.


    \begin{figure}[t]
    \includegraphics[width=1.0\linewidth]{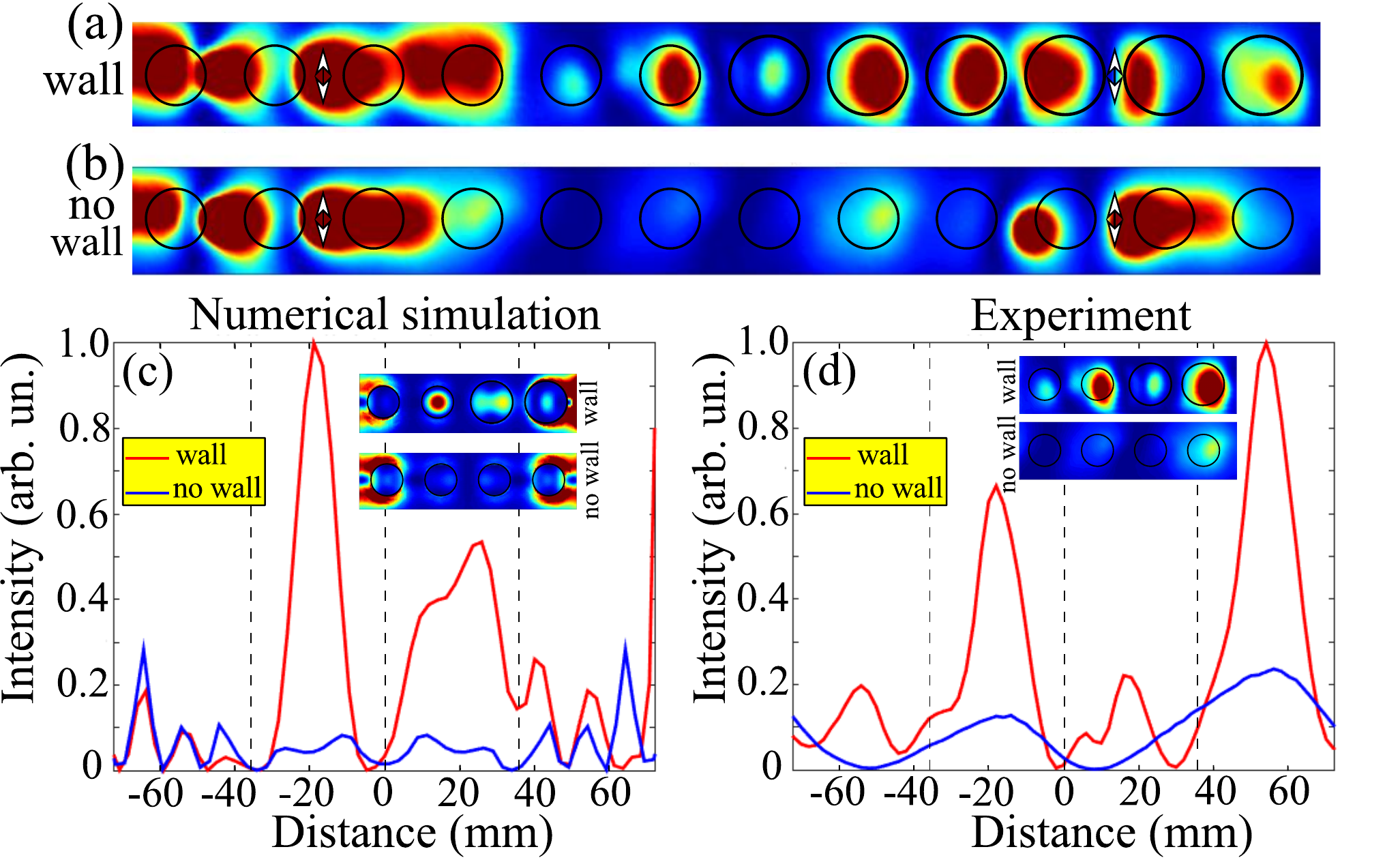}
    \caption{Experimental observation of the topological interface state excited by the pair of electric dipoles in the array of 12 bianisotropic disks at frequency $f=2.435$~GHz. (a,b) Experimental near field maps showing the quantity $I=|E_x|^2+|E_y|^2$ for the array (a) with and (b) without the domain wall. White arrows show the location of the dipoles exciting the structure. (c,d) Normalized intensity $I$ along the axis of the array according to (c) numerical simulation in CST Microwave Studio software package and (d) experimental data. Distance is measured from the array center. Insets illustrate field distribution in the vicinity of array center for the cases with and without domain wall.} 
    \label{fig:Experiment}
    \end{figure}

{\it Experiment}~--~To confirm our prediction of Jackiw-Rebbi-type interface states, we have designed and tested experimentally a microwave prototype system based on high-index ceramic meta-atoms purchased from ``Ceramics Co. Ltd.'' having real and imaginary parts of permittivity equal to $\eps'=39$ and $\eps''=0.004$, respectively. Each of meta-atoms is composed of two coaxial cylinders with different diameters ($d_1=29.1$~mm, $d_2=22.0$~mm) and different heights ($h_1=9.0$~mm and $h_2=3.0$~mm). The cylinders are placed on top of each other as illustrated in Fig.~\ref{fig:Array}(a,b). Such design breaks the meta-atom mirror symmetry with respect to $Oxy$ plane which gives rise to magneto-electric coupling. Parameters of cylinders were chosen in order to ensure that electric and magnetic dipole resonances appear in the same frequency range around $2.4-2.5$~GHz.

The fabricated 12 ceramic particles were arranged in a linear array with the period $R=36$~mm. The meta-atoms were placed into the holes drilled in an extruded polystyrene matrix with the permittivity close to 1 in the frequency range of interest. The structure was excited by two electric dipoles placed at two edges of the structure and separated by eight meta-atoms. The dipoles were connected to the port of vector network analyzer (VNA) Rohde \& Schwarz ZVB20 through the divider which provided in-phase feeding. Near field measurement was accomplished using automatic mechanical high precision scanner Trim TMC 3113 and subwavelength electric dipole connected to the second port of VNA which was used as the receiving antenna. To suppress multiple reflections, the measurements were carried out in an anechoic chamber with the walls covered by microwave absorbers Eccosorb VHP-12-NRL.

The measurements of the near field map were performed for the two structures: first structure with the same alignment of meta-atoms (``no wall" case) and the second one with half of the meta-atoms in the array flipped over, which ensured the opposite signs of magneto-electric coupling $\mu$ for the two halves of the array [``wall'' case, Fig.~\ref{fig:Array}(a)]. In agreement with our theoretical predictions, we observe the interface states (Fig.~\ref{fig:Experiment}) which manifest themselves through field localization in the vicinity of the domain wall.

Furthermore, our simulations suggest that the observed interface states are quite robust to disorder, whereas buckling of the array into zigzag geometry is accompanied by the topological transition from Jackiw-Rebbi interface states to the topological edge states (see Supplementary Materials for details).

{\it Discussion and conclusions}~--~To summarize, we have demonstrated an approach to tailor coupling constants in effective photonic Hamiltonians by {\it local} modification of the disk bianisotropy. We believe that this finding bridges the existing gap between the need for tunable nonlinear couplings required for implementation of tunable  topological states~\cite{Hadad} from one side and experimental possibilities of nonlinear tuning allowing only for modification of on-site parameters. While at microwave frequencies this problem is typically solved by insertion of varactor diodes~\cite{Hadad-Nature,Dobrykh,Serra-Garcia-prb}, the implementation of the same functionality in infrared or optical range remains highly challenging.

Besides that, we envison that our findings will initiate further research on photonic topological states in metamaterials with the staggered bianisotropy pattern which can be readily implemented experimentally but still waits for detailed theoretical investigations. We hope that our work makes the first step in this exciting direction.


{\it Acknowledgments}~--~We acknowledge valuable discussions with Dmitry Filonov. This work was supported by the Russian Foundation for Basic Research (Grant No.~18-32-20065). Experimental studies were partially supported by RFBR grant No.~19-02-00939 and by the Ministry of Education and Science of the Russian Federation (Zadanie No.~3.2465.2017/4.6). M.G. acknowledges partial support by the Foundation for the Advancement of Theoretical Physics and Mathematics ``Basis''.

\bibliography{TopologicalLib}

\end{document}